# A unified N-SECE strategy for highly coupled piezoelectric energy scavengers

**A Morel[1,2], A Badel[2], Y Wanderoild[1] and G Pillonnet[1]**

[1] Univ. Grenoble Alpes, CEA, LETI, MINATEC, F-38000 Grenoble, France
[2] Univ. Savoie Mont Blanc, SYMME, F-74000 Annecy, France

Email: adrien.morel@cea.fr

**Abstract**. This paper proposes a novel vibration energy harvesting strategy based on an extension of the Synchronous Electric Charge Extraction (SECE) approach, enabling both the maximization of the harvested power and a consequent bandwidth enlargement in the case of highly coupled/lightly damped piezoelectric energy harvesters. The proposed strategy relies on the tuning of the frequency of the energy extraction events, which is either $N$ times greater than the vibration frequency (*Multiple* SECE case, $N > 1$) or $1/N$ times smaller (*Regenerative* SECE, $N < 1$). We first prove analytically than increasing or decreasing $N$ both lead to a damping reduction. While $N$ has no impact on the system's resonance frequency in the *Regenerative* case ($N < 1$), we show that this resonant frequency becomes a function of $N$ in the *Multiple* SECE case ($N > 1$). Experimental results on a highly coupled/lowly damped piezoelectric harvester ($k^2 = 0.44, Q_m = 20$) demonstrates the potential of this strategy, leading to 257% harvested power improvement compared to SECE ($N = 1$). and the possibility to tune the resonant frequency on a range as large as 35% of the short-circuit resonant frequency of the harvester.

## 1. Introduction

Energy scavenging is a promising alternative to batteries in order to supply sensor nodes. Vibrations energy harvesting is of particular interest in closed confined environments where there are few solar energies nor thermal gradients. At a cm scale, piezoelectric energy harvesters (PEH) are a good way to convert mechanical strain into useful and storable electrical energy, since they are easier to miniaturize than electromagnetic generators, while exhibiting higher power densities than electrostatic generators [1].

During the last decade, electrical interfaces for this kind of harvester have been studied. Standard interfaces based on a diode bridge rectifying and smothering the piezoelectric voltage have first been proposed in [2,3]. However, these interfaces, in the case of lowly coupled PEH, only allow to extract a small part of the maximal energy that could be harvested [4]. Furthermore, their efficiencies are highly dependent on the voltage of the storage capacitance.

In order to design higher performances interfaces, researchers have developed SECE based strategies [4]. Indeed, they propose simple, efficient, and load independent ways to harvest the energy from lowly coupled and/or highly damped PEH. These strategies have either be implemented using on-the-shelf components [5], or developing dedicated ASIC with external passive components [6].

Thanks to recent progresses in the mechanical design and the use of highly coupled piezoelectric materials, e.g. single crystal, piezoelectric harvesters may now exhibit much higher electromechanical coupling associated with low mechanical damping. For this particular kind of PEH, SECE strategy is not well adapted since it overdamps the mechanical resonator, leading to low displacements and harvested energy [7,8].

Some research has been conducted to investigate new efficient strategies that could be used for highly coupled PEH, in order to optimize the damping induced by the electrical interface [8-10]. Indeed, in [8], we proposed to wait a certain number of semi-period before harvesting the accumulated energy in order to reduce the damping induced by the electrical interface. In [9] and [10], the damping is reduced thanks to the control of the number of electrical charges extracted from the piezoelectric material. Those two approaches, even though they allowed to reduce the damping, did not induce any frequency tuning effect which could although be used to enlarge the harvesting bandwidth.

Electrically-based frequency tuning strategies have also been proposed in the literature [11-15]. In [11], a capacitive bank combined with a variable resistive load emulated thanks to a DC/DC converter allows to tune the phase of the piezoelectric voltage while optimizing the damping induced by the interface. However, this approach requires an important number of passive components. In order to overcome this issue, in [12], we proposed to replace the capacitive bank with a short-circuit control which emulates a tunable capacitive behavior. [13] and [14-15] proposed a new frequency tuning approach based on the combination of the tunable SECE explained in [9] and [10] with a phase tuning. This approach is theoretically very efficient, however it introduces two tuning parameters, leading to a complicated control scheme which may take time to converge. Since the harvested energy may be a complex function of these two parameters, it may also converge toward a low local maximum, leading to limited performances.

Recently, researchers have started to propose single tuning harvesting strategies, which could combine damping optimization, some frequency tuning possibilities, with a simpler single dimension control scheme. In [16], Lefeuvre et al proposed a phase-shifted SECE strategy which consists of only controlling the phase between the displacement extrema and the energy extraction event. This strategy allows to reach the maximum harvestable power for two different frequencies with a single tuning parameter. However, this technique is quite sensitive to the phase, requiring a sensitive control loop. Plus, the two frequencies for which the harvested power is maximized are both functions of the coupling coefficient, which tend to complicate the design of applications-dedicated PEHs.

In this paper, we propose a new single tuning harvesting approach. Based on the strategy presented in [8], we generalize the regenerative principal for any case where the energy extraction events happen at a different frequency compared to the vibration frequency. We prove that there is two different ways to reduce the damping, either by harvesting the energy at a lower frequency rate, as detailed in [8], or at a higher frequency rate, which is a new and complementary way to reduce the damping. Interestingly, the energy extraction frequency has indeed a high impact on the dynamics of the electromechanical system, leading to frequency tuning possibilities. This new strategy, while exhibiting equivalent performances as the one presented in [16], is more robust and less sensitive to the variations of the tuning parameter. Indeed, good performance in terms of maximal harvested power and bandwidth can be obtained with only 2 or 3 different values of $N$. Furthermore, one of the frequency where the harvested power is maximized is insensitive to the electromechanical coupling, which greatly facilitates the PEH design for this kind of approach.

In a first part, we remind why the standard SECE becomes less efficient when the electromechanical coupling of the harvester increases. In a second part, we present the proposed approach called N-SECE strategy. We then define two different cases ($N < 1$ and $N > 1$) which are theoretically studied. The power expressions in each case are derived, and lead to a unified analysis of the influences of $N$ on the dynamics and performance of the PEH. Finally, the proposed approach is experimentally validated, and the obtained results are described and discussed.

## 2. Original background: standard SECE

The standard SECE has first been introduced by Lefeuvre et al. in [4]. It is an interesting alternative to the standard energy harvesting (SEH) interface based on a full bridge rectifier, for two main reasons:

- It increases the percentage of mechanical energy of the harvester converted into electrical energy, leading to enhanced performances on lowly coupled and/or highly damped PEH.
- Its performances are not dependent on the voltage across the storage element ($V_{dc}$ in Figure 1).

The SECE strategy can be described as follow: during a semi period, the energy is accumulated in the piezoelectric material. When the voltage reaches its maximum value, an inductance $L$ is connected to the piezoelectric material. The electrical energy stored on the piezoelectric element is then transferred to the inductance. Once the voltage on the piezoelectric element reaches zero, the inductance is disconnected from the piezoelectric element and the energy flows to the storage element $C_{out}$. The typical circuit realizing the SECE strategy and its associated waveforms are shown in Figure 1 and Figure 2.

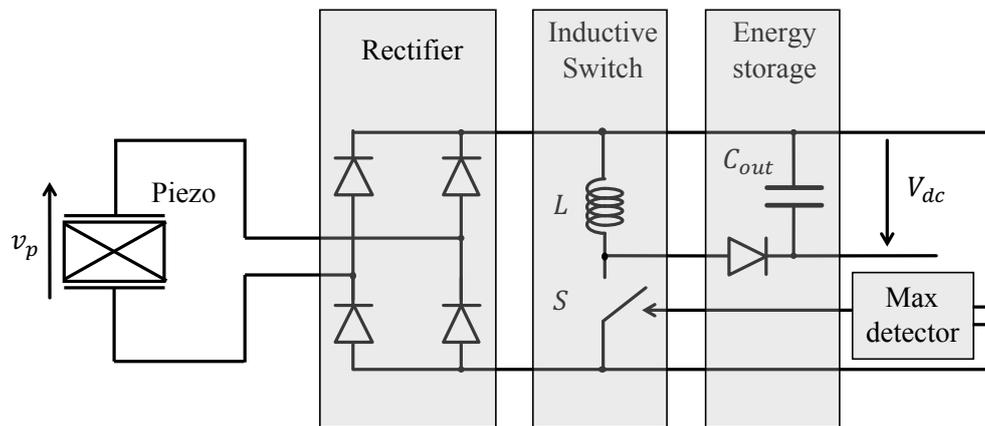

Figure 1-SECE interface circuit

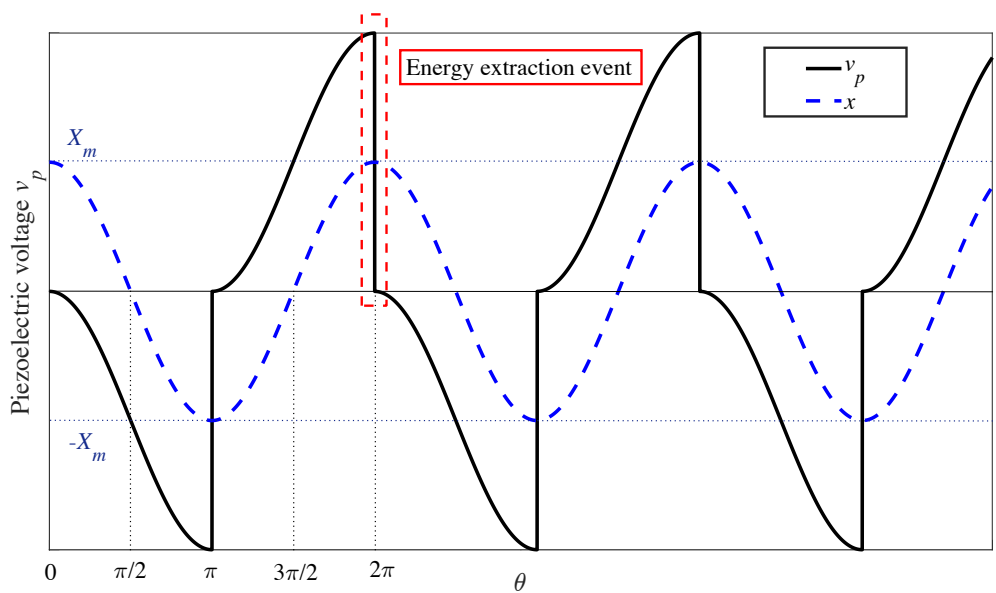

Figure 2-Typical voltage and displacement waveforms when using the SECE strategy.

In order to theoretically evaluate SECE performance, we will consider a PEH made of a piezoelectric material deposited on a mechanical linear oscillator, as shown in Figure 3.

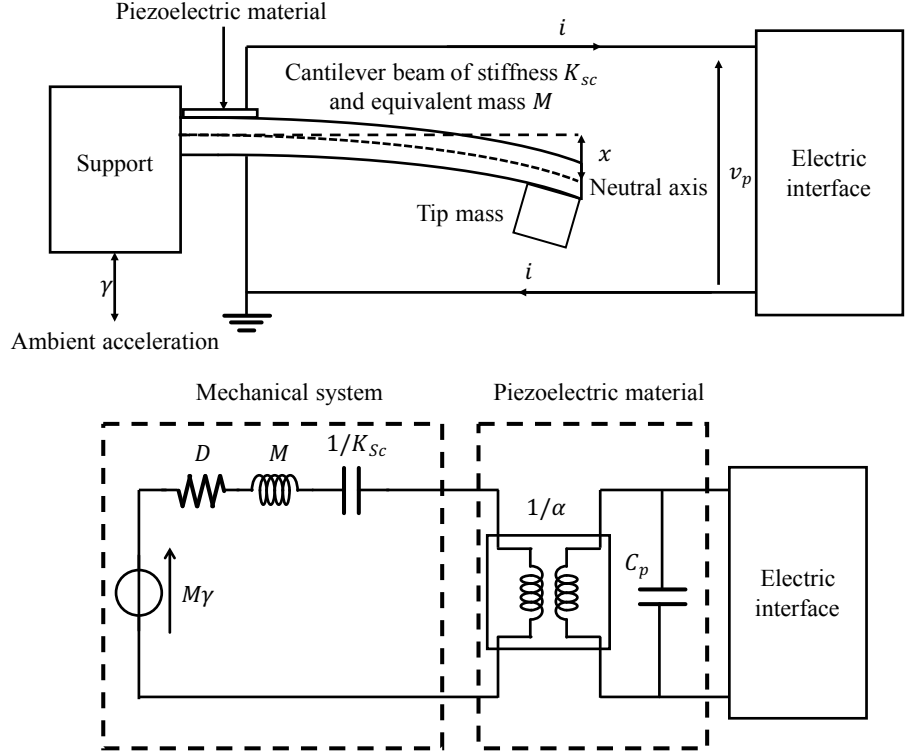

Figure 3-Piezoelectric energy harvester made of a cantilever beam, and its equivalent electrical model

This PEH can be seen as a single-degree of freedom mass + damper + stiffness + piezo system, that is classically modelled by the equivalent electrical model shown in figure 3 and whose constitutive equations are given by (1).

$$\begin{cases} F = M\gamma = M\ddot{x} + D\dot{x} + K_{sc}x + \alpha v_p \\ \quad\quad\quad i = \alpha\dot{x} - C_p\dot{v}_p \end{cases} \quad (1)$$

Where $F$, $\gamma$, $x$, $i$ and $v_p$ respectively stand for the vibration force, ambient acceleration, the displacement of the inertial mass, the current extracted in the interface circuit, and the voltage across the electrodes of the piezoelectric material. $M$, $D$, $K_{sc}$, $C_p$ and $\alpha$ stand for the dynamic mass of the harvester, the mechanical damping, the short-circuit stiffness, the piezoelectric material clamped capacitance, and the piezoelectric coefficient.

This model is accurate as long as the harvester remains linear and that the PEH is driven around one of its resonant frequencies. In order to simplify the calculations, we will consider in the following that $x$ is purely sinusoidal and of constant amplitude, as expressed by (2). This physically means that the harvester is considered in steady state operation and that any potential mechanical non-linearity is neglected.

$$x(t) = X_m \cos(\theta) = X_m \cos(\omega t) \quad (2)$$

As explained extensively in [17], the studies of the electrical interfaces for piezoelectric harvesters can be unified and simplified thanks to normalized variables:

$$\begin{cases} k_m^2 = \dfrac{k^2}{1-k^2} = \dfrac{\alpha^2}{K_{sc}C_p} \\ Q_m = \dfrac{\sqrt{K_{sc}M}}{D} \\ \omega_0 = \sqrt{\dfrac{K_{sc}}{M}} \\ \Omega_m = \omega/\omega_0 \\ \omega_{oc} = \omega_0\sqrt{1+k_m^2} \end{cases} \quad (3)$$

$k^2 \in [0,1]$ is the square coupling coefficient, while $k_m^2 \in \mathbb{R}^+$ is defined as the square modified electromechanical coupling. $Q_m, \omega_0$ and $\Omega_m$ are the quality factor of the mechanical resonator, its short circuit angular resonant frequency, and the normalized vibration frequency respectively. The two normalized variables $k_m^2$ and $Q_m$ are sufficient to characterize and predict the performances of a linear PEH, and their product $k_m^2 Q_m$ is commonly used to evaluate the potential of a piezoelectric harvester [8,17]. That is why it has been defined as a figure of merit in previous works [17]. $\omega_{oc}$ is the open circuit angular resonant frequency of the harvester. If the PEH is highly coupled, this frequency is slightly different than the short circuit angular resonant frequency, $\omega_0$.

From (1) and (2) and considering the diode threshold voltages negligible compared to the piezoelectric voltage, it can be proven that (for a SECE interface) the displacement amplitude $X_m$ can be given by the following expression when the vibration frequency matches the resonance frequency of the PEH:

$$X_m = \frac{F_m}{D\omega + \dfrac{4\alpha^2}{\pi C_p}} \quad (4)$$

Where $F_m$ stands for the amplitude of the vibration force $F$. The harvested power thanks to the SECE interface is given by (5).

$$P = \frac{2\omega\alpha^2}{\pi C_p} \frac{F_m^2}{\left(D\omega + \dfrac{4\alpha^2}{\pi C_p}\right)^2} \approx \frac{16 k_m^2 Q_m}{\pi} \frac{P_{max}}{\left(1 + \dfrac{4 k_m^2 Q_m}{\pi}\right)^2} \quad (5)$$

Where $P_{max}$ is the maximum power which can be extracted from a linear PEH under a constant vibration of amplitude $F_m$. The expression of this maximum power is given by (6).

$$P_{max} = \frac{F_m^2}{8D} = \frac{M\gamma_m^2}{8\omega_0} Q_m \quad (6)$$

Where $\gamma_m$ is the amplitude of the driving acceleration. Figure 4 represents the evolution of the harvested power at resonance with both the SECE and the standard interface as a function of the product $k_m^2 Q_m$. We can clearly see that for $k_m^2 Q_m = \pi/4$, the harvested power is maximized. When the product $k_m^2 Q_m$ gets too important, the SECE overdamps the mechanical system, leading to a reduction of the vibration amplitude and lower performances than what can be obtained with a simple rectifier. This can also be seen in Figure 5, which shows the displacement amplitude at the resonant frequency as a function of $k_m^2 Q_m$. When the PEH is too highly coupled and/or too lightly damped, the displacement amplitude goes below its optimal value, leading to a lower extracted energy [7,8].

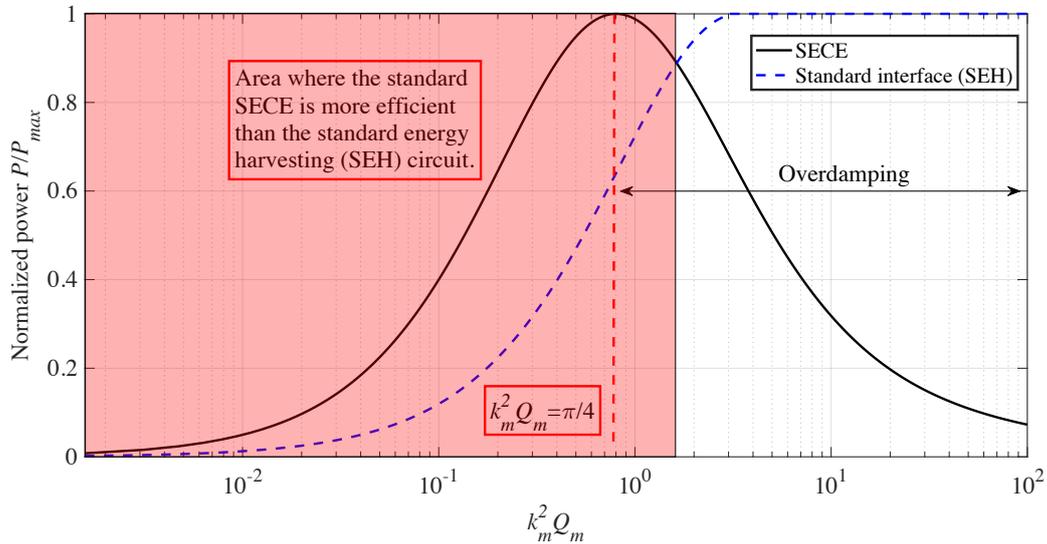

Figure 4- Normalized harvested power using SECE (black) and standard interface (blue) as a function of $k_m^2 Q_m$.

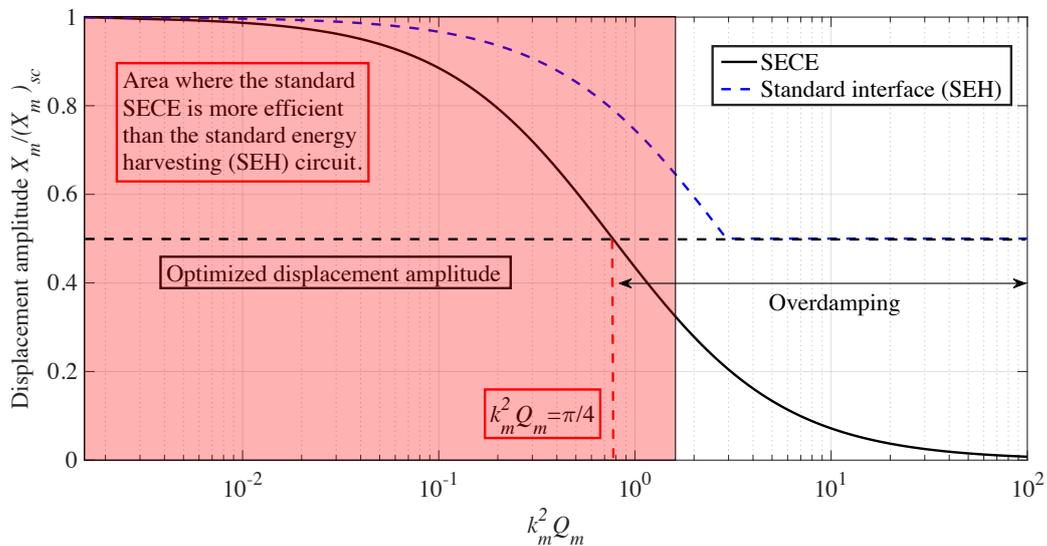

Figure 5-Normalized displacement amplitude using SECE (black) and standard interface (blue) at resonance as a function of $k_m^2 Q_m$.

$(X_m)_{sc}$ given by $F_m/(D\omega_0)$ is the short circuit amplitude displacement of the PEH at resonance. This is the maximum value that can take the displacement, as there is no damping induced by the electrical interface when the piezoelectric material electrodes are short circuited.

Because of this overdamping phenomenon, we should find ways to reduce the damping induced by the SECE interface in order to be able to use it on high coupling interfaces. This could lead to an interface:
- Which is effective on any PEH, whether it is highly coupled/damped or not.
- Whose effectiveness is not related to the voltage across the storage element.

The SECE strategy increasing the forces induced by the electrical interface on the mechanical system, we could use this characteristic to electrically change the resonant frequency of the electromechanical system, and enlarge the harvesting energy bandwidth.

### 3. Proposed concept: N-SECE

In the standard SECE, there are two voltage extrema every vibration's period, thus the SECE harvesting frequency is two times higher than the vibration frequency ($f_{harv_{SECE}} = 2f_{vib}$). In the proposed N-SECE, instead of harvesting the energy at a frequency fixed by the vibrations, similarly as the standard SECE strategy, we propose to harvest the energy at a different frequency, controlled by the parameter $N$, where $N=f_{harv}/f_{harv_{SECE}} = f_{harv}/(2f_{vib})$ as depicted in Figure 6.

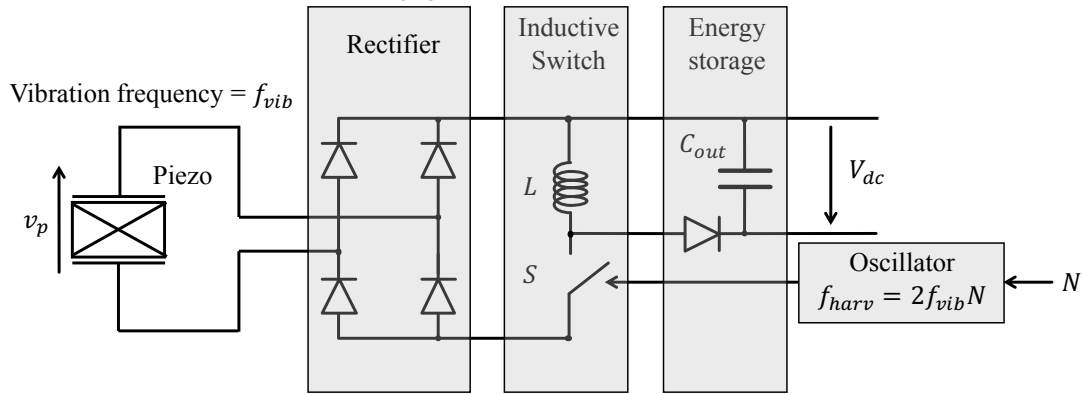

Figure 6-Proposed N-SECE system based on a tunable parameter $N$

By doing so, we obtain the piezoelectric voltage waveforms shown in Figure 7 when the harvesting frequency $f_{harv}$ is higher than $2f_{vib}$ ($N > 1$). We also obtain the voltage waveforms shown in Figure 8 when $f_{harv}$ is lower than $2f_{vib}$ ($N < 1$). Obviously, when $N = 1$, $f_{harv} = f_{harv_{SECE}}$, and the N-SECE interface behaves like a standard SECE interface. Increasing or decreasing $N$ leads in any case to a reduction of the damping induced by the electrical interface, and hence to improved performance, as demonstrated in the next sections.

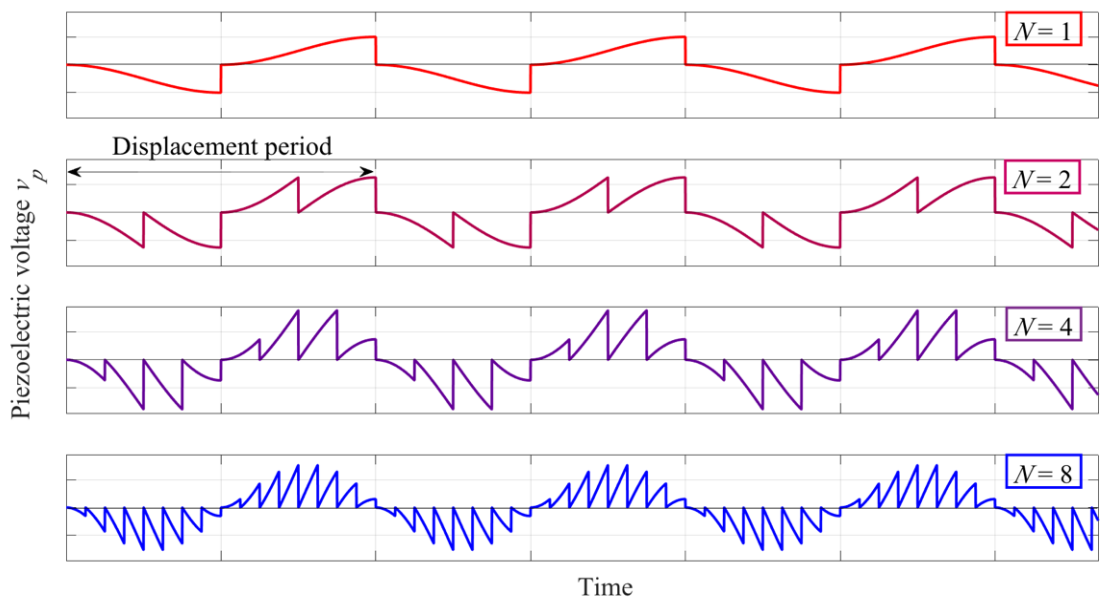

Figure 7-Piezoelectric voltage waveforms for various $N$ when $N > 1$

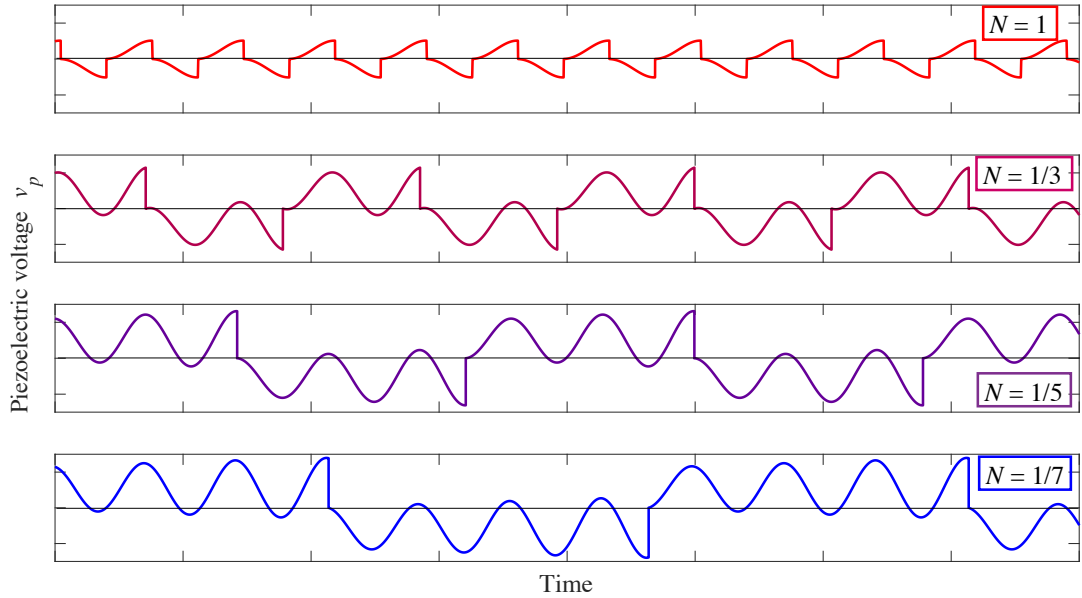

Figure 8-Piezoelectric voltage waveforms for various $N$ when $N < 1$

## 4. Theoretical analysis: *Multiple SECE* (*N>1* case)

In this section, the case where the harvesting frequency is higher than the vibration frequency ($N > 1$) is considered. In this case, $N \in \mathbb{N}^*$ and $n \in [1, N] \cap \mathbb{N}^*$ respectively represent the number of harvesting events in a vibration semi-period, and an index of the harvesting event in the semi-period. First, the PEH works in open circuit condition: the voltage across the piezoelectric material electrodes starts to increase as charges accumulate in the dielectric capacitance. During every harvesting event, this piezoelectric voltage drops almost instantly to zero as the electrical charges are extracted from the material. Then the voltage starts to grow again, until the next harvesting event. Consequently, the piezoelectric voltage during a semi-period can be divided in $N$ subset voltages, each related to a particular harvesting event. The expression of the $n^{th}$ subset voltage is given by (7)

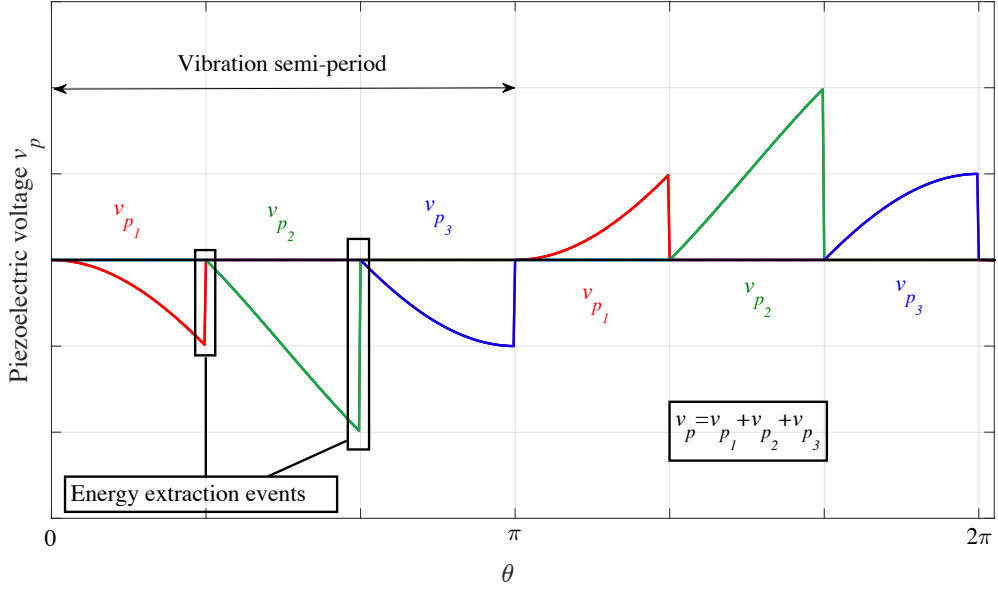

Figure 9-Theoretical voltage waveforms obtained for $N = 3$, revealing 3 different subset voltages

$$v_{p_n}(\theta) = \begin{cases} \dfrac{\alpha}{C_p} \int_{\frac{(n-1)}{N}\pi}^{\theta} \dot{x}(\theta)d\theta, & \forall \theta \in ]\dfrac{(n-1)}{N}\pi, \dfrac{n\pi}{N}[ \\ 0, & \forall \theta \in [0, \dfrac{(n-1)}{N}\pi] \cup [\dfrac{n\pi}{N}, \pi] \end{cases} \quad (7)$$

Consequently, the piezoelectric voltage expression during a semi-period is the sum of $n$ periodic subset voltages, and can be expressed as a Fourier series since it is a periodical signal.

$$v_p(\theta) = \sum_{n=1}^{N} v_{p_n}(\theta) = \sum_{n=1}^{N} \sum_{u=1}^{\infty} [a_{u_n} \cos(u\theta) + b_{u_n} \sin(u\theta)] \quad (8)$$

Figure 7 shows the piezoelectric voltage waveforms simulated for various $N$ while Figure 9 illustrates the case $N = 3$. In order to linearize the piezoelectric voltage expression given by (8), we find the first harmonic coefficients of the Fourier series associated with $v_p$. These coefficients are given by (9).

$$\begin{cases} a_1 = \dfrac{X_m \alpha}{2C_p \pi} \sum_{n=1}^{N} \left[ \sin\left(\dfrac{2n\pi}{N}\right) + \sin\left(\dfrac{2(n-1)\pi}{N}\right) + \dfrac{2\pi}{N} - 2\cos(\dfrac{(n-1)\pi}{N})\sin(\dfrac{n\pi}{N}) \right] \\ b_1 = \dfrac{X_m \alpha}{2C_p \pi} \sum_{n=1}^{N} \left[ -\cos\left(\dfrac{2n\pi}{N}\right) + \cos\left(\dfrac{2(n-1)\pi}{N}\right) + 4\cos(\dfrac{(n-1)\pi}{N})(\cos(\dfrac{n\pi}{N}) - \cos(\dfrac{(n-1)\pi}{N})) \right] \end{cases} \quad (9)$$

Considering the filtering effect of the mechanical resonator, it is reasonable to consider that only the first voltage harmonic $v_{p_1}$ will have an influence on the dynamics of the system. Its expression in the Fourier domain $\underline{v_{p_1}}$ is given by (10).

$$\underline{v_{p_1}}(\theta) = \left(\dfrac{a_1}{X_m} - j\dfrac{b_1}{X_m}\right)\underline{x}(\theta) \quad (10)$$

Where $\underline{x}$ stands for the mass displacement written in the Fourier domain. Substituting (10) in (1) expressed in the Fourier domain leads to the expression of the displacement magnitude $X_m$, given by (11).

$$\begin{cases} X_m = \dfrac{F}{\sqrt{A^2 + B^2}} \\ A = K_{sc} - M\omega^2 + \alpha \dfrac{a_1}{X_m} \\ B = \omega D - \alpha \dfrac{b_1}{X_m} \end{cases} \quad (11)$$

From the displacement amplitude obtained in (11), we can eventually determine the harvested power, given by (12).

$$P = \frac{\alpha^2}{C_p} X_m^2 \frac{\omega}{2\pi} \sum_{n=1}^{N} \left[ \cos\left(\frac{n\pi}{N}\right) - \cos\left(\frac{(n-1)\pi}{N}\right) \right]^2 \quad (12)$$

Thus, from (9), (11) and (12), the harvested power can be determined for any value of $N$ as a function of the PEH's parameters and of the ambient acceleration frequency and magnitude.

## 5. Theoretical analysis: *Regenerative SECE* (*N*<1 case)

When the harvesting frequency $f_{harv}$ is lower than the vibration frequency $f_{vib}$, the voltage waveforms is different, leading to a completely different analytical formulation, fully detailed in [8]. In this case, $N$ can be defined as the inverse of $P \in \mathbb{N}^*$, where $P$ represents the number of semi-period of vibrations between each harvesting events. The piezoelectric voltage is then expressed by (13).

$$v_p(\theta) = \begin{cases} \dfrac{1}{C_p} \displaystyle\int_0^{\theta/N} \alpha \dot{x}(\theta) d\theta, & \forall \theta \in\, ]0, \pi/N[ \\ 0, \ if\ \theta = \pi/N \\ \dfrac{1}{C_p} \displaystyle\int_\pi^{\theta/N} \alpha \dot{x}(\theta) d\theta, & \forall \theta \in\, ]\pi/N, 2\pi/N[ \\ 0, \ if\ \theta = 2\pi/N \end{cases} \quad (13)$$

This piezoelectric voltage has been drawn numerically for various values of $N$, as depicted in Figure 8. The new voltage expression leads to another Fourier expression. The expressions of the first Fourier coefficients are given by (14).

$$\begin{cases} a_1 = \dfrac{\alpha X_m}{C_p} \\ b_1 = 2 \dfrac{\alpha X_m N}{C_p \pi} \left((-1)^{1/N} - 1\right) \end{cases} \quad (14)$$

Thus, in the *regenerative* case (*N*<1), as extensively explained in [8], the displacement amplitude (15) as well as the harvested power expressions (16) can be derived. Note that when $1/N$ is even, the additional damping due to energy harvesting is null, and thus the harvesting power is zero. We will only consider odd values of $1/N$ in the following.

$$\begin{cases} X_m = \dfrac{F}{\sqrt{A^2 + B^2}} \\ A = K_{sc} - M\omega^2 + \dfrac{\alpha^2}{C_p} \\ B = \omega D - 2\dfrac{\alpha^2 N}{C_p \pi}\left((-1)^{1/N} - 1\right) \end{cases} \qquad (15)$$

$$P = \dfrac{N\omega\alpha^2}{2\pi C_p} X_m^2 \left((-1)^{1/N} - 1\right)^2 \qquad (16)$$

## 6. Results and discussions

Based on the first harmonic analysis, two analytical models for the harvesting power in the two possible cases of the proposed approach are derived in the previous sections ($N < 1$ and $N > 1$). We will now analyze these results in order to understand their physical insights. From the precedent part, we can determine the expressions of the stiffness $K_{elec}$ and of the damping $D_{elec}$ induced by the electrical interface as two functions of $N$ for both cases ($N < 1$ and $N > 1$), as shown in Table 1. The stiffness $K_{elec}$ has a direct impact on the angular resonant frequency $\omega_{res}$ of the harvester, given by (17). The damping $D_{elec}$ in another hand, impacts the electromechanical quality factor $Q_{em}$ of the system, given by (18).

Table 1 – Influences of $N$ on the stiffness and damping induced by the electrical interface

| | $N \leq 1$ | $N > 1$ |
|---|---|---|
| $(K)_{elec}$ | $\dfrac{\alpha^2}{C_p}$ | $\dfrac{\alpha^2}{2\pi C_p} \sum_{n=1}^{N}\left[\sin\left(\dfrac{2n\pi}{N}\right) + \sin\left(\dfrac{2(n-1)\pi}{N}\right) + \dfrac{2\pi}{N} - 2\cos\left(\dfrac{(n-1)\pi}{N}\right)\sin\left(\dfrac{n\pi}{N}\right)\right]$ |
| $(D)_{elec}$ | $-N\dfrac{2\alpha^2}{C_p\pi}\left((-1)^{1/N} - 1\right)$ | $\dfrac{\alpha^2}{C_p\pi}\sum_{n=1}^{N}\left[-\cos\left(\dfrac{2n\pi}{N}\right) + \cos\left(\dfrac{2(n-1)\pi}{N}\right) + 4\cos\left(\dfrac{(n-1)\pi}{N}\right)\left(\cos\left(\dfrac{n\pi}{N}\right) - \cos\left(\dfrac{(n-1)\pi}{N}\right)\right)\right]$ |

$$\omega_{res} = \sqrt{\dfrac{(K_{sc} + K_{elec})}{M}} \qquad (17)$$

$$Q_{em} = \dfrac{\sqrt{M(K_{sc} + K_{elec})}}{D + D_{elec}} \qquad (18)$$

Figure 10 shows the damping induced by the electrical interface normalized with respect to the damping induced by a standard SECE interface as a function of $N$ in the case of a highly coupled PEH ($k_m^2 = 0.8$). Figure 11 shows the electromechanical resonant frequency of the same PEH as a function of $N$. When $N < 1$ (*regenerative SECE* case), decreasing $N$ reduces the damping induced by the electrical interface on the mechanical part of the system. However, the frequency corresponding to the maximum harvested power (the resonant frequency) is not modified by $N$. Interestingly, when $N > 1$ (*multiple SECE* case), increasing $N$ not only reduces the damping due to the electrical interface, but also tune the resonant frequency of the electromechanical system, which converges toward the short-circuit resonant frequency of the system. When $N = 1$, the two expressions given by Table 1 become the same, which confirms the coherence of the two models and that the two cases ($N < 1$ and $N > 1$) can be gathered in a single unified one ($N > 0$).

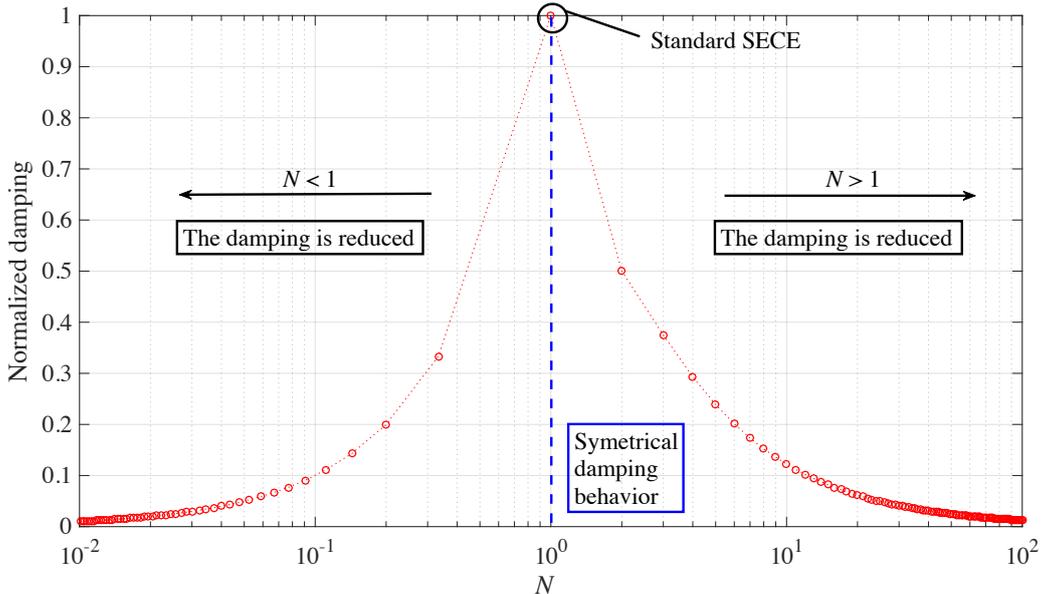

Figure 10-Normalized damping induced by the N-SECE interface as a function of $N$ ($k_m^2 = 0.8, Q_m = 20$).

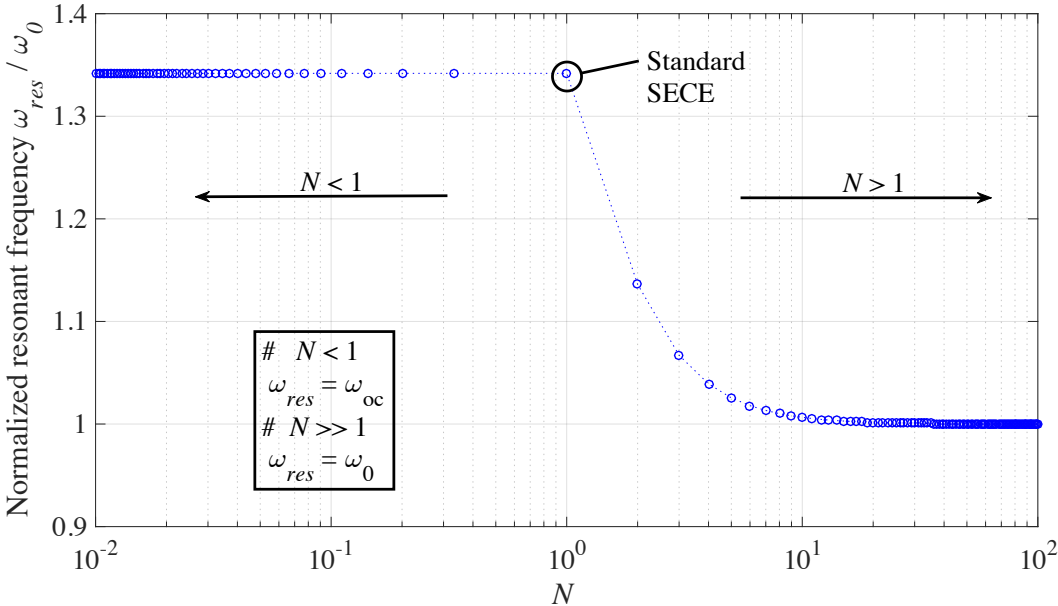

Figure 11-Normalized electromechanical resonant frequency as a function of $N$ ($k_m^2 = 0.8, Q_m = 20$).

Some power frequency responses are illustrated for various $N$ in Figure 12. As shown in Figure 12, tuning $N$ allows to reach the maximum harvestable energy for two different frequencies: one corresponding to the short-circuit resonant frequency of the system $\omega_0$, and the other one being close to the open circuit resonant frequency $\omega_{OC}$. Indeed, it is well known that for any piezoelectric harvester, there exists an optimal damping induced by the electrical interface which maximizes the harvested power [17, 18]. Figure 10 shows that this optimal damping can either be reached by increasing $N$ (*multiple SECE*) or decreasing $N$ (*regenerative SECE*), but these two actions lead to two different resonant frequencies since the voltage waveforms are different and do not influence the system's dynamics in the same way. The dashed line green envelope in Figure 12 shows the maximum harvestable power with the N-SECE interface. The associated $N_{opt}$ are shown in Figure 13.

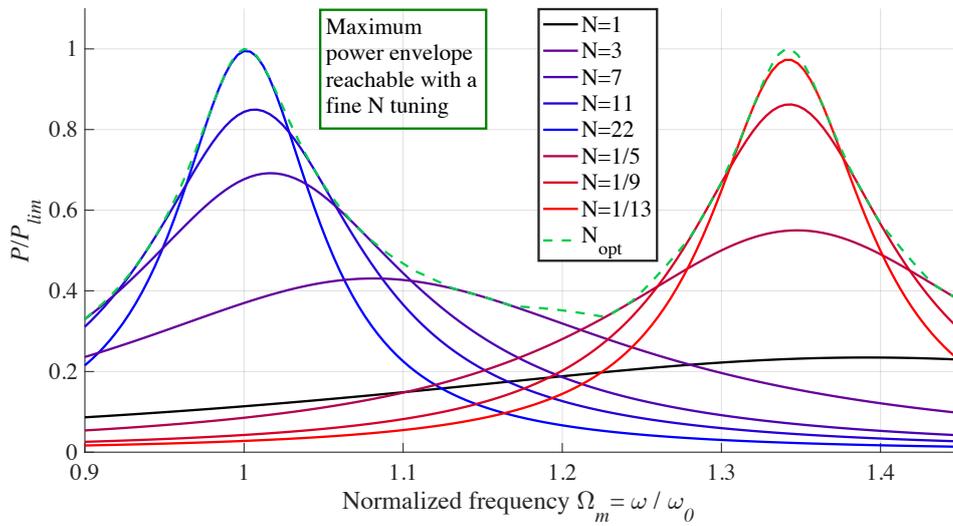

Figure 12-Power frequency responses of the N-SECE strategy for various $N$ ($k_m^2 = 0.8, Q_m = 20$).

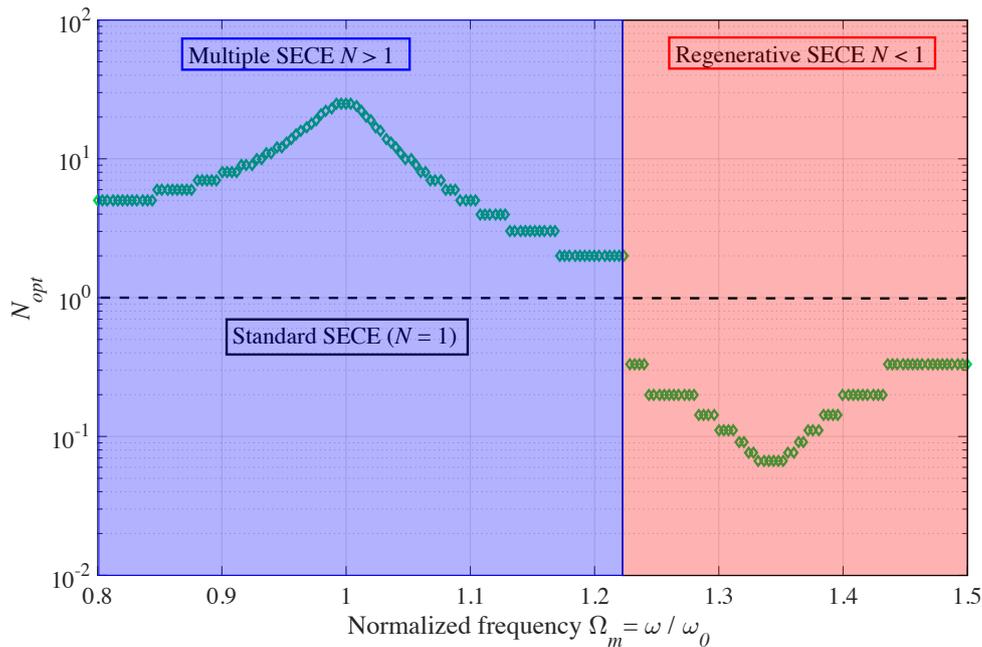

Figure 13-Optimal $N$ as a function of the normalized frequency $\Omega_m$ leading to the green envelope shown in Figure 12.

We propose in the following to compare the performances of the N-SECE with the SECE and other single parameter tuning strategies, i.e. the *tunable* SECE [9] and the *phase-shifted* SECE [16]. The behaviours of these strategies with different vibration frequencies as well as with different PEHs' figure of merit $FoM_{PEH}$ have been studied. Thus, in the following part, the harvestable energy areas in the $[k_m^2 Q_m, \Omega_m]$ plane are shown, described, and analysed.

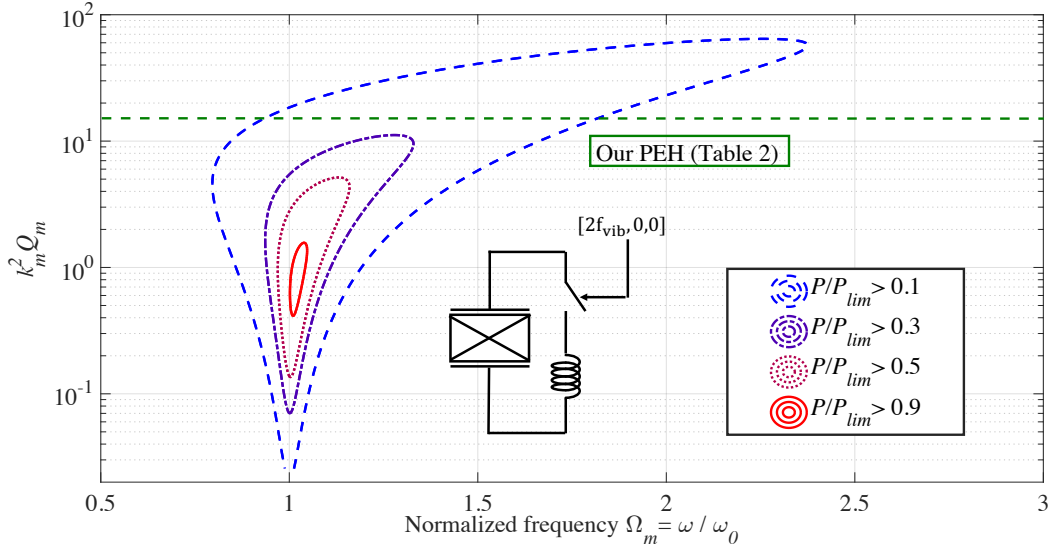

Figure 14-Harvestable power in the $[k_m^2 Q_m, \Omega_m]$ plane using a SECE interface [4], with $Q_m$=20

As explained previously and as shown in Figure 14, the standard SECE is only optimized at one particular frequency, at a particular coupling. The Figure 14 shows how this strategy is particularly inefficient for highly coupled piezoelectric harvesters. In order to overcome the overdamping of SECE, some strategies have already been proposed in the state of the art. All the proposed strategies, including the one presented in this paper, are based on a tuning of the energy extraction events, which can be characterized by a set of three parameters: $[f_{harv}, \phi_{harv}, \beta]$. $f_{harv}$ is the frequency of the energy extraction events, $\phi_{harv}$ is the phase between the displacement extrema and the energy extraction events, and $\beta$ accounts for the voltage remaining in the piezoelectric capacitance $C_p$ after the energy extraction event, and can be defined as $\beta = v_p(t_0^+)/v_p(t_0^-)$ with $t_0$ being the switching instant. In the case of the standard SECE, the switching law is ruled by the following parameters:

$$\begin{cases} f_{harv} = 2f_{vib} \\ \phi_{harv} = 0 \\ \beta = 0 \end{cases} \quad (19)$$

The *tunable* SECE was previously reported by Lefeuvre et al. in [9], and consists in controlling the damping induced by the electrical interface thanks to the parameter $\beta \in [0,1]$. This parameter has thus directly an influence on the quantity of charges extracted from the piezoelectric material. The new set of switching parameters are expressed by (20), while the harvested power in the $[k_m^2 Q_m, \Omega_m]$ plane can be seen in Figure 15. The parameter β has been optimized for each point of Figure 15.

$$\begin{cases} f_{harv} = 2f_{vib} \\ \phi_{harv} = 0 \\ \beta \in [0,1] \end{cases} \qquad (20)$$

As seen in Figure 15, the frequency where the harvested power is maximized is highly dependent on the coupling of the harvester. Thus, it is somehow important to take into account the coupling of the harvester while designing the mechanical structure and choosing its resonant frequency, which may be complicated.

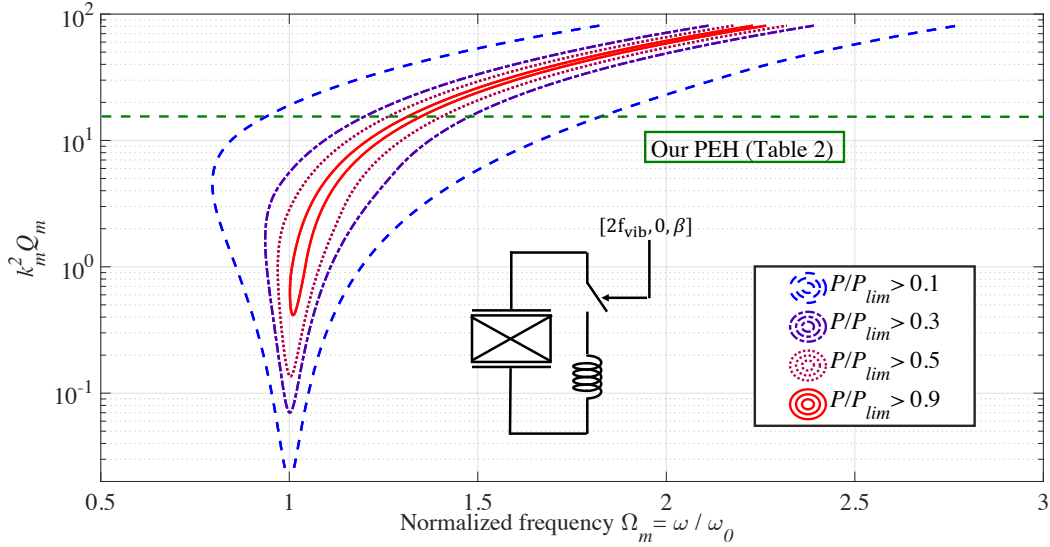

Figure 15-Harvestable power in the $[k_m^2 Q_m, \Omega_m]$ plane using a tunable SECE interface [9], with $Q_m = 20$

Another proposed strategy, the *phase-shifted* SECE, consists in controlling the moment where the energy extraction events occur [16]. This is done thanks to a phase $\phi_{harv} \in [-\pi/2, \pi/2]$ between the energy extraction events and the displacement extremum. The switching parameters then become:

$$\begin{cases} f_{harv} = 2f_{vib} \\ \phi_{harv} \in [-\pi/2, \pi/2] \\ \beta = 0 \end{cases} \qquad (21)$$

The harvested power in the $[k_m^2 Q_m, \Omega_m]$ plane can be observed in Figure 16. Each point has been determined with an optimal phase $\phi_{harv}$ maximizing the harvested power. This strategy exhibits interesting characteristics, as shown in Figure 16, as there are two frequencies for which the harvested power can be maximized. This leads to an improved harvesting bandwidth compared to the *tunable* SECE. However, these two frequencies are both dependent on the coupling of the harvester. Furthermore, the phase of the harvesting event is hard to accurately control, because of the intrinsic phase-shift already induced by the electronic control loop [19,20].

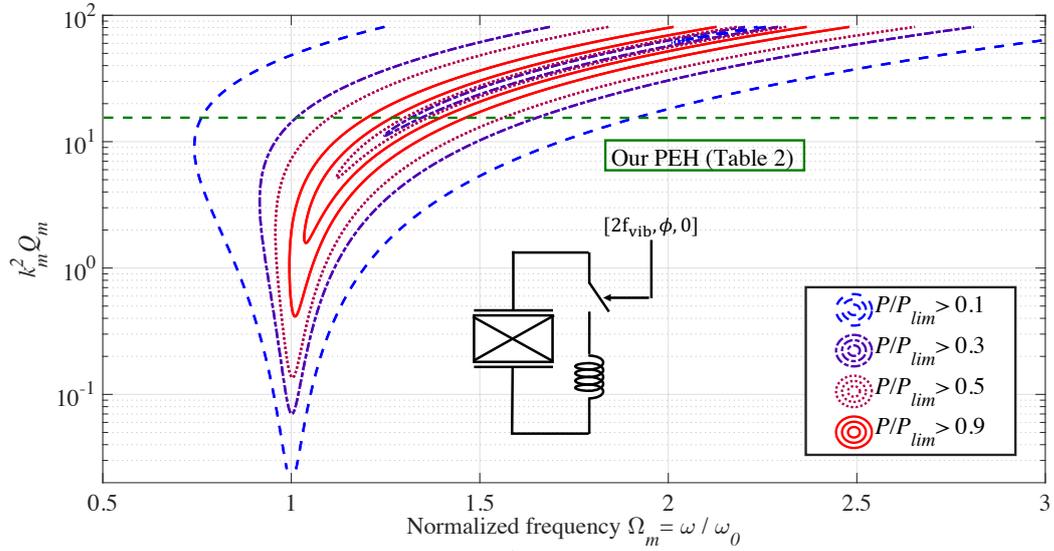

Figure 16-Harvestable power in the $[k_m^2 Q_m, \Omega_m]$ plane using a phase-shifted SECE interface [16], with $Q_m = 20$

Finally, the strategy proposed here is based on the tuning of a parameter $N$, which can be seen as a control of the energy extraction events frequency. Thus, the switching parameters can be expressed as:

$$\begin{cases} f_{harv} = 2Nf_{vib} \\ \phi_{harv} = 0 \\ \beta = 0 \end{cases} \quad (22)$$

The harvested power thanks to the N-SECE can be observed in Figure 17. Each point has been determined with an optimal N maximizing the harvested power. As shown in Figure 17, the N-SECE also exhibits two frequencies for which the harvested power is maximized, leading to a better bandwidth than the *tunable* SECE strategy. One of this frequency is not dependent on the coupling of the PEH, which is a considerable advantage. Moreover, as it will be shown in the experimental part, the N-SECE allows to have good performances with only a few $N$ values.

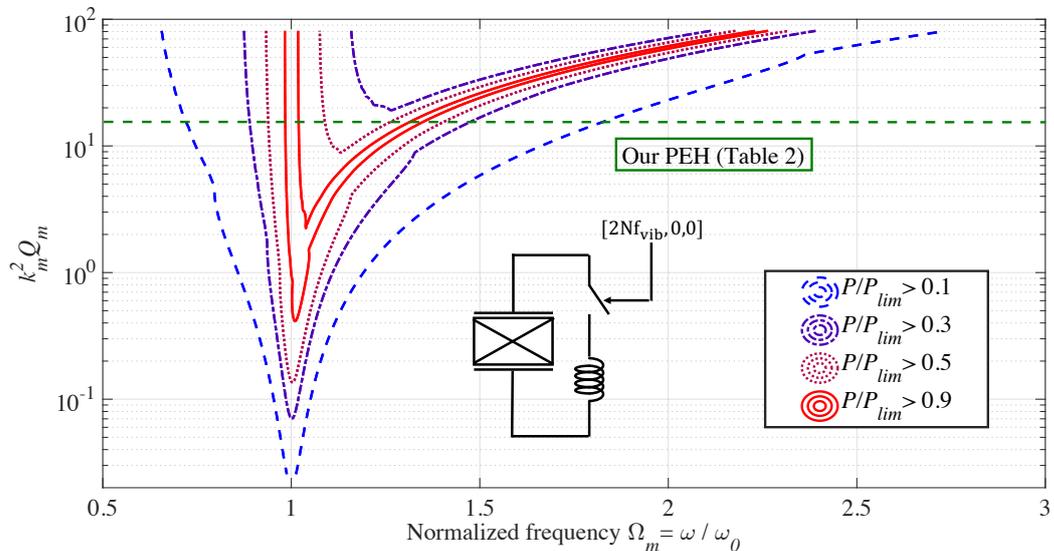

Figure 17-Harvestable power in the $[k_m^2 Q_m, \Omega_m]$ plane using the N-SECE, with $Q_m = 20$

Figure 18 summarizes the power responses of our harvester with state of the art interfaces (SECE [4], *tunable* SECE [9], *phase-shifted* SECE [16]), the proposed N-SECE, and the standard electrical harvesting interface SEH whose diode bridge's output voltage has been optimized for every point of Figure 18. Due to the high coupling and low damping of our harvester ($k_m^2 Q_m = 16$), the SECE overdamps the system, as explained previously.

The standard interface allows theoretically to get relatively good harvesting performance. However, to reach the power response shown in black in Figure 18, the output voltage of the diode bridge rectifier should be dynamically adapted. This is usually done thanks to a DC/DC converter whose input voltage and impedance are functions of its switching signal's duty cycle, its frequency, and of the output voltage of the converter, which is as well the voltage across the storage element [2, 21]. This command seems difficult to realize, because the optimal input voltage value depends on the vibration amplitude, and on the vibration frequency. To the authors knowledge, a switching law allowing such optimal control cannot be found in prior art. Furthermore, the input impedance of the DC/DC converter should be adapted from approximately $200\Omega$ to $1M\Omega$ in order to reach the performances of Figure 18. The several decades between these two impedances make the design of such command complicated.

The *tunable* SECE, *phase-shifted* SECE, and the proposed N-SECE greatly enhance the harvested power compared to the SECE. It can be noted that their effectiveness doesn't depend on the vibration amplitude, nor on the voltage across the storage element. Therefore, it seems easier to implement those strategies in real case applications compared to the standard electrical harvesting interface.

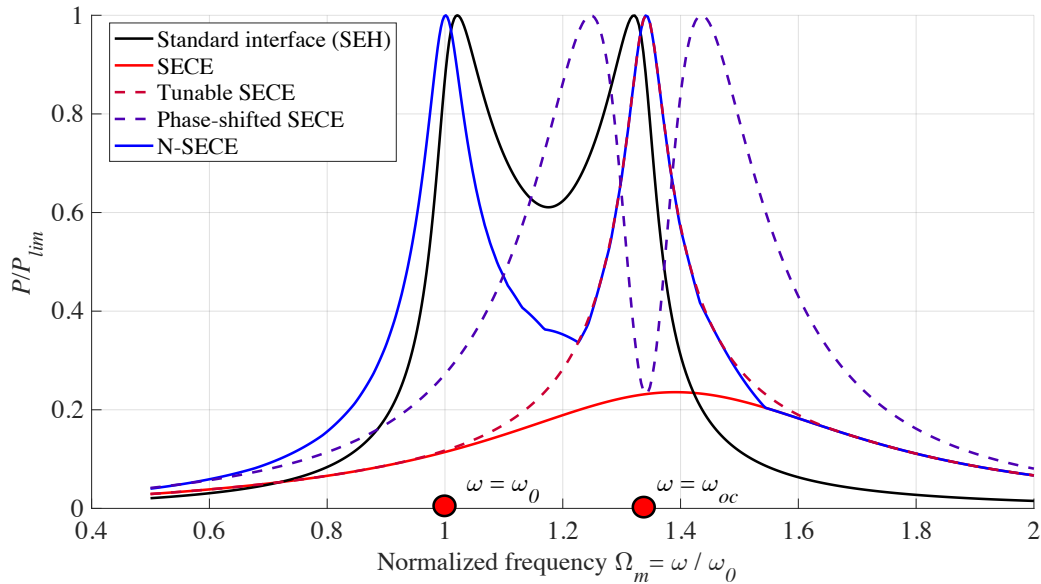

Figure 18-Simulated power responses of our harvester (Table 2) with various electrical interfaces.

Finally, the results of this part are summarized in Figure 19. It shows the $FoM$ defined by (23) for any harvesting strategy, which quantifies how much a strategy is better than the standard SECE on the whole frequency range. As we can see, SECE-based strategies are the most efficient for low coupling harvesters. As the coupling is increased, the SECE's $FoM$ remains constant while other strategies' $FoM$ increase. From Figure 19, several conclusions can be drawn:

- When the product $k_m^2 Q_m$ is inferior or equal to 1, the SECE-based strategies are more efficient than the standard electrical harvesting interface (SEH).
- When the product $k_m^2 Q_m$ is inferior or equal to 1, it is not interesting to tune the SECE strategy, as all the SECE-based strategies share the same $FoM$.
- When the product $k_m^2 Q_m$ is greater than 1, any tuning increases the $FoM$ compared to the SECE. The *tunable* SECE strategy is less efficient than the *phase-shifted* SECE and the N-SECE because it cannot be used to tune the electromechanical resonant frequency of the PEH. Indeed, as shown on Figure 18, its power peak is centered around $\omega_{oc}$.
- For any $k_m^2 Q_m$ the *phase-shifted* SECE and N-SECE have a better $FoM$ than the standard interface.

Overall, the best strategies, as expected, are the *phase-shifted* SECE and the N-SECE since they both exhibit two power peaks. In practice, the fact that one of the N-SECE power peak is not dependent on the PEH's coupling is a strong advantage. Furthermore, as shown in the experimental part, the N-SECE allows to reach good harvested power on a relatively large frequency band with only a few $N$ values. This makes the N-SECE, in our opinion, the best choice among all the tuning strategies.

$$FoM = \frac{\int_0^\infty P_{strategy} d\omega}{\int_0^\infty P_{SECE} d\omega} \qquad (23)$$

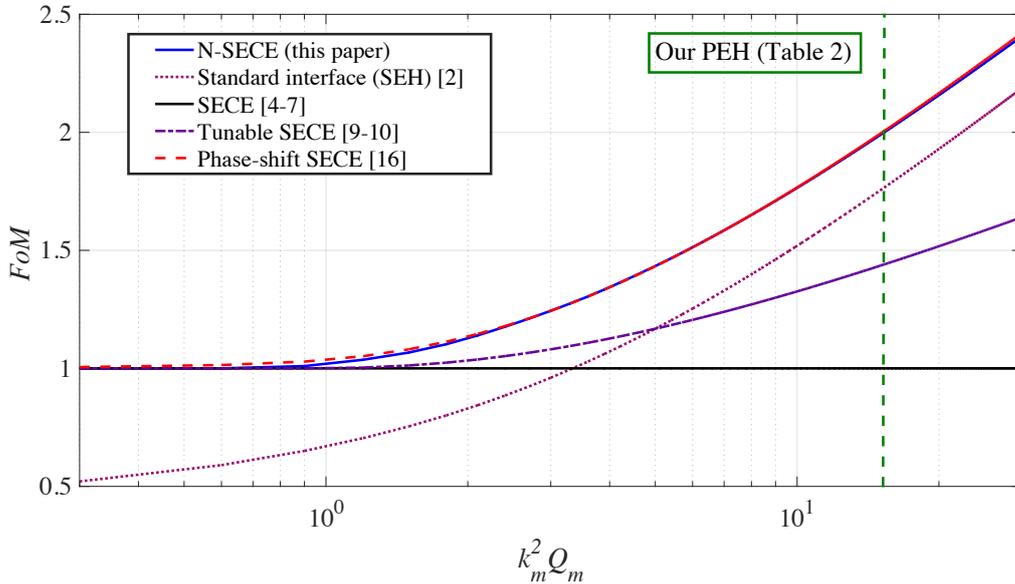

Figure 19-FOM of the different tuning strategies as functions of $k_m^2 Q_m$ with $Q_m = 20$

## 7. Experimental results

In order to validate the proposed theoretical model, we ran experimental tests on a highly coupled piezoelectric harvester. The experimental setup is schematically illustrated in Figure 20. An electromagnetic shaker is driven to generate a sinusoidal constant amplitude ($|\gamma| = 0.3 m.s^{-2}$) acceleration whose frequency can be controlled. The piezoelectric harvester is fixed on the shaker, and its speed, displacement, and voltage waveforms are recorded for any couple of parameters $(\omega, N)$. A N-SECE interface circuit made of off-the-shelf discrete components has been realized in order to enable the harvesting event frequency to be tuned. The whole setup has been automated thanks to simulink and a matlab script through a Dspace real-time control system. The experimental tests have been realized for 200 different values of $\omega$, and 20 different values of $N$.

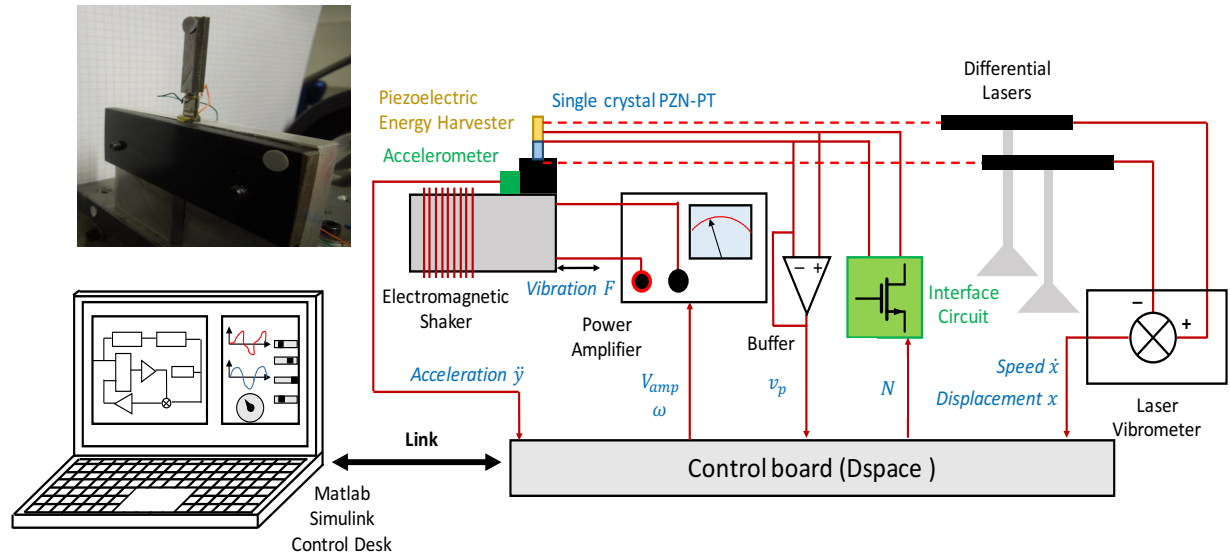

Figure 20-Automated test bench for the N-SECE strategy experimental verification

The highly coupled PEH is composed of two $10\times5\times0.5$ mm³ single crystal PZN-PT plates deposited on a cantilever beam. This PEH characteristics are summarized in Table 2.

Table 2 – PEH parameters obtained under an acceleration of $|\gamma| = 0.3 m.s^{-2}$

| Parameters names | Parameters values | Units |
|---|---|---|
| $k^2$ | 0.44 | - |
| $k_m^2$ | 0.8 | - |
| $Q_m$ | 20 | - |
| $P_{max}$ | 7.2 | µW |
| $C_p$ | 1.4 | nF |
| $\omega_0$ | 612.4 | rad.s$^{-1}$ |

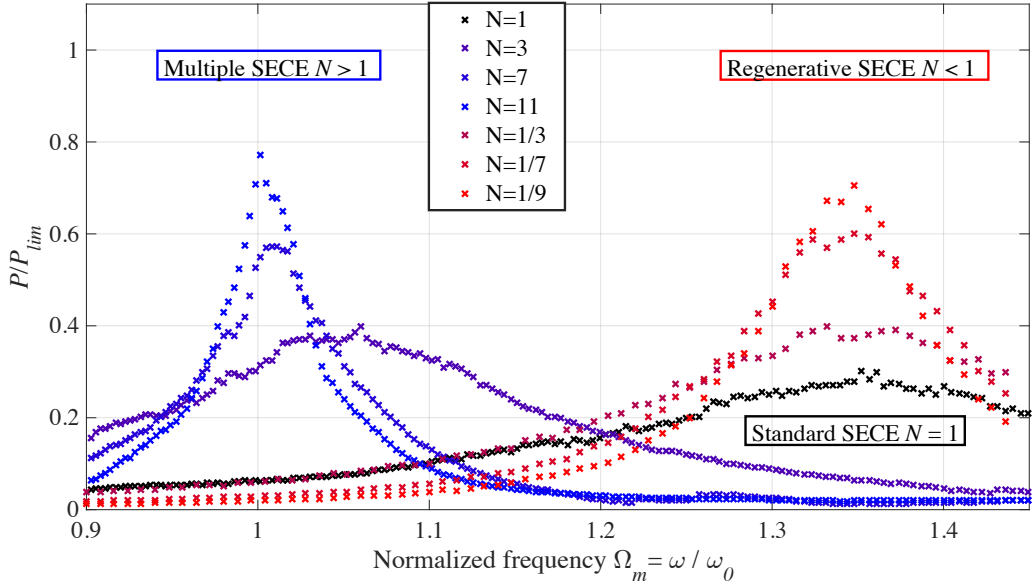

Figure 21-Experimental power frequency responses of the proposed N-SECE strategy when $|\gamma| = 0.3 m.s^{-2}$

The experimental power frequency responses of the proposed N-SECE strategy under a very low acceleration amplitude ($|\gamma| = 0.3 m.s^{-2}$) are shown on Figure 21. This low acceleration value has been set because of the voltage limitations of our electrical interface. Indeed, the high coupling of our PEH (important $\alpha$, low $C_p$) leads to high piezoelectric voltages (At resonance, when $|\gamma| = 0.3 m.s^{-2}$, the amplitude of the open circuit piezoelectric voltage is around $10V$) even when the acceleration is small. These results can be extended for higher vibration accelerations if the PEH's capacitance $C_p$ is increased, or by choosing higher voltage rating electrical components. The results have been normalized using (3) and (6) in order to be consistent with the theoretical results presented in Figure 12. The tendencies predicted by the theoretical model and shown in Figure 12 are in good agreement with the experimental results. The expected behaviours, i.e. the increase of power for both the *multiple* SECE and *regenerative* SECE, as well as the frequency tuning effect when $N > 1$ have been experimentally verified. The differences and the unpredicted energy losses may be attributed to the dielectric losses [22] and the higher resonance modes of the mechanical resonator which may get excited due to the non-linear processing of the piezoelectric voltage.

The power spectrum presents two power peaks. The first one reaches $5.7 \mu W$ at the short circuit resonance frequency of our PEH, 97.5Hz. The second one corresponds to the open circuit resonance frequency of our PEH, 130.8Hz, and allows to harvest $5.2 \mu W$. These two peaks are both greatly improving the harvested power compared to the standard SECE (black dots on Figure 21) of about 20 times and 2.6 times respectively. With $N$ taking only two values {1/3,3}, we are able to harvest more than 20% of $P_{max}$ ($1.44 \mu W$) on a large frequency band starting from 90Hz to 145Hz, which corresponds to 56% of the short circuit frequency of the PEH. These results clearly prove that an adequate tuning of $N$ leads to important power harvesting enhancement compared to the standard SECE (black crosses in Figure 21).

Practical implementation of the N-SECE is similar to previous implementations of the SECE strategy which has been successfully realized using on-the-shelf components [5] or with a dedicated ASIC with external passive components [6, 23]. A way to control the frequency of the energy extraction events is however required. This could be done using a low frequency oscillator whose oscillating frequency would be determined by a controller, as shown in Figure 22.

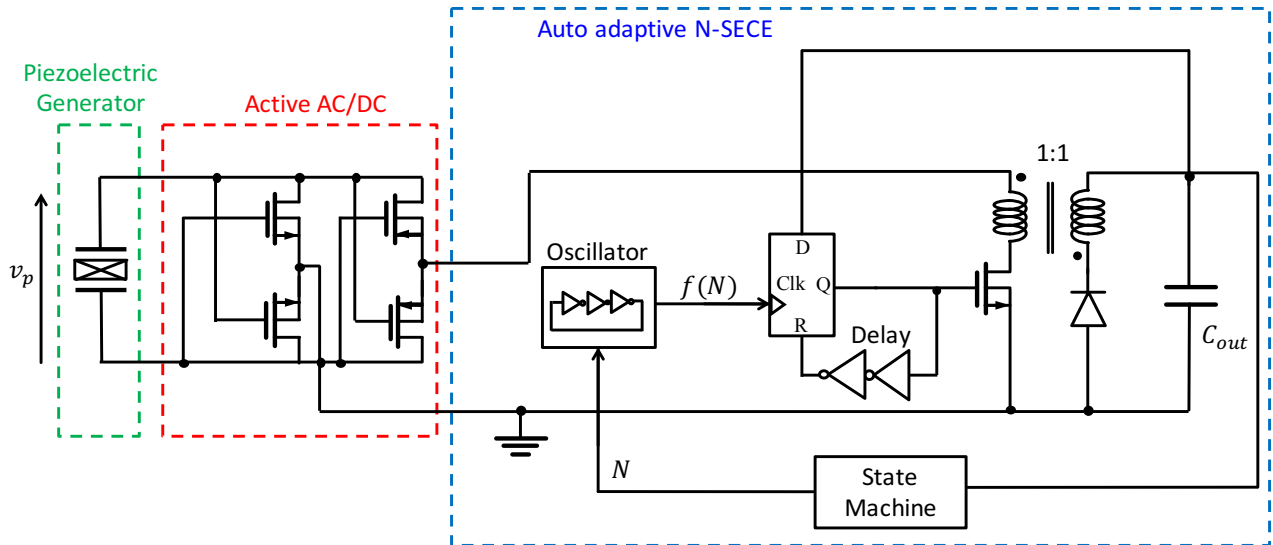

Figure 22-Proposed implementation of the N-SECE

In order to reach the power frequency responses envelope drawn in green on Figure 12 we will need to implement as well an ultra-low power control on $N$, represented in Figure 22 by the "State Machine" block. This block could be for instance a gradient algorithm optimizing the energy going to the storage element $C_{out}$. A few predefined and wisely chosen $N$ values (1/3 and 3 for example) could guarantee the fast convergence and low power consumption of the algorithm while ensuring that the harvested power would remain high on a relatively large frequency band.

## 8. Conclusion

In this paper, we present a new non-linear harvesting strategy for PEH characterized by high values of $k_m^2 Q_m$. After introducing the linear PEH electromechanical model, we reminded how standard SECE is no more efficient when the product $k_m^2 Q_m$ is superior to unity. We then proposed an analytical modelling of the N-SECE, a novel strategy which drastically enhances the harvested power and bandwidth when $k_m^2 Q_m > 1$. The performances of this strategy (power, frequency bandwidth) has been analyzed both analytically and experimentally. It allows an improvement with our PEH of the maximal harvested power of up to 257% compared to the standard SECE and more than $1.44\mu W$ (20% of $P_{max}$) on a frequency band from 90Hz to 145Hz under very small vibration amplitude ($\gamma = 0.3 m.s^{-2}$) with only two different $N$ (1/3 and 3). Around the short-circuit resonant frequency of the harvester $\omega_0$ this approach even allows to harvest more than 20 times more power than the standard SECE. The fact that it allows both a frequency tuning and a power optimization is promising and simple interface, switching dynamically (thanks to an ultra-low power MPPT algorithm) between 2 or 3 values of $N$, optimizing the harvested energy on a relatively large frequency band has to be developed.

## 9. Acknowledgments

This work was supported by the French Alternative Energies and Atomic Energy Commission.